\documentclass[a4paper,12pt]{article}

\usepackage{amssymb}
\usepackage{latexsym}
\usepackage{amsmath}

\usepackage[dvips]{graphicx}

\usepackage{times}
\usepackage[T1]{fontenc}
\normalfont
\usepackage{mathptmx}

\usepackage{stmaryrd}

\usepackage{vmargin}
\setpapersize{A4}

\usepackage{mathrsfs}

\def\B2S{\mathop{\rm B2S} \nolimits}

\newcommand{\MBB}{\mathbb}

\newcommand{\VEC}{\overrightarrow}

\newcommand{\TENS}{\otimes}

\def\linearimp{\mathop{- \hspace{-.025in} \circ}}

\newcommand{\LIMP}{\linearimp}

\newtheorem{proposition}{Proposition}[section]

\begin{document}
\title{A New Proof of P-time Completeness of Linear Lambda Calculus}
\author{National Institute of Advanced Industrial Science and Technology (AIST), \\
1-1-1 Umezono, Tsukuba, Ibaraki, 305-8563 Japan \\
 \\ Satoshi Matsuoka}

\maketitle
\begin{abstract}
We give a new proof of P-time completeness of 
Linear Lambda Calculus, which was originally given by H. Mairson in 2003. 
Our proof uses an essentially different Boolean type from the type Mairson used. \\
Moreover the correctness of our proof can be machined-checked using an implementation of Standard ML. 
\end{abstract}

\section{Introduction}
In \cite{Mai04}, H. Mairson gave a proof of P-time completeness of Linear Lambda Calculus. 
It is an excellent exercise of linear functional programming.
The crucial point of the proof is that the copy function of truth values is representable by a linear term:
this is relatively easy in Affine Lambda Calculus as shown in \cite{Mai04}, but
quite difficult in Linear Lambda Calculus. 
So, the key issue there is to avoid the use of the weakening rule. 
The issue was also treated from a different angle in \cite{Mat07}, which 
established typed B\"{o}hm theorem without the weakening rule. \\
In this paper we give a new proof of P-time completeness of Linear Lambda Calculus.
Our proof is different from that of \cite{Mai04} in the following points: 
\begin{itemize}
\item In \cite{Mai04} Mairson used the base Boolean type 
$\MBB{B}_{\rm MH} = p \LIMP p \LIMP (p \LIMP p \LIMP p) \LIMP p$ while 
we use $\MBB{B} = (p \LIMP p) \LIMP (p \LIMP p) \TENS (p \LIMP p)$. 
Although both have two normal forms, 
they are different because 
while $\MBB{B}_{\rm MH}$ reduces to itself by 
the {\it linear distributive transformation} given in Section 3 of \cite{Mat07}
(which was called third order reduction in \cite{Mat07}), 
$\MBB{B}$ reduces to 
\[ \MBB{B}_{\rm red} = p \LIMP p \LIMP (p \LIMP p) \LIMP (p \LIMP p \LIMP p) \LIMP p, \]
which has six normal forms. 
\item All the two variable functions that can be representable over $\MBB{B}_{\rm MH}$ {\it without any polymorphism} are only {\it exclusive or} and its negation, but 
in $\MBB{B}$ they are all the boolean functions except for exclusive or and its negation,
i.e., fourteen functions. 
\item our proof is also an interesting application of the linear distributive transformation. 
\end{itemize}
As in \cite{Mai04}, our proof is also machine-checkable:
all the linear $\lambda$-terms in this paper are also well-formed expressions of Standard ML \cite{MTHM97}.
So the reader may confirm the correctness of our proof using an implementation of Standard ML.
We used the interactive system of Standard ML of New Jersey.

\section{Typing Rules}
We give our term assignment system for Linear Lambda Calculus.
Our system is based on Natural Deduction, e.g., given in \cite{Tro92},
which is equivalent to the system based on Sequent Calculus or proof nets in \cite{Gir87}
(see \cite{Tro92}). 
Its notation is unusual in the Linear Logic community, but
its purpose is to make our proof machine-checkable. 
\paragraph{Types}
\[
\mbox{\tt A} ::= \mbox{\tt 'a} \, \, \, \, | \, \, \, \, \mbox{\tt A1*A2} \, \, \, \, | \, \, \, \, \mbox{\tt A1->A2}
\]
The symbol {\tt 'a} stands for a type variable.
On the other hand $\mbox{\tt A1*A2}$ stands for the tensor product $\mbox{\tt A1} \TENS \mbox{\tt A2}$ and $\mbox{\tt A1->A2}$ for the linear implication $\mbox{\tt A1} \LIMP \mbox{\tt A2}$.
\paragraph{Terms}
We use {\tt x,y,z} for term variables, 
$\VEC{\mbox{\tt x}}, \VEC{\mbox{\tt y}}, \VEC{\mbox{\tt z}}$ for 
finite lists of term variables,
and {\tt t,s} for general terms.

\paragraph{Term Assignment System}
\[
\frac{}{\mbox{\tt x:A} \vdash \mbox{\tt x:A}}
\]
\[
\frac{\mbox{\tt x:A}, \VEC{\mbox{\tt y}} \mbox{\tt :} \Gamma \vdash \mbox{\tt t:B}}
{\VEC{\mbox{\tt y}} \mbox{\tt :} \Gamma \vdash \mbox{\tt fn x=>t:A->B}}
\quad \quad 
\frac
{ \VEC{\mbox{\tt x}} \mbox{\tt :} \Gamma \vdash \mbox{\tt t} : \mbox{\tt A->B} \quad \VEC{\mbox{\tt y}} \mbox{\tt :} \Delta \vdash \mbox{\tt s:A} }
{ \VEC{\mbox{\tt x}} \mbox{\tt :} \Gamma, \VEC{\mbox{\tt y}} : \Delta \vdash \mbox{\tt ts:B}}
\]
\[
\frac
{ \VEC{\mbox{\tt x}} \mbox{\tt :} \Gamma \vdash \mbox{\tt s:A} \quad \VEC{\mbox{\tt y}} \mbox{\tt :} \Delta \vdash \mbox{\tt s:B} }
{ \VEC{\mbox{\tt x}} \mbox{\tt :} \Gamma, \VEC{\mbox{\tt y}} : \Delta \vdash \mbox{\tt (s,t):A*B}}
\quad \quad 
\frac
{ \VEC{\mbox{\tt x}} \mbox{\tt :} \Gamma \vdash \mbox{\tt s:A*B}
\quad \mbox{\tt x:A}, \mbox{\tt y:B}, \VEC{\mbox{\tt y}} \mbox{\tt :} \Delta \vdash \mbox{\tt t:C} }
{ \VEC{\mbox{\tt x}} \mbox{\tt :} \Gamma, \VEC{\mbox{\tt y}} \mbox{\tt :} \Delta \vdash \mbox{\tt let val (x,y)=s in t end:C} }
\]
Moreover, function declaration \\
\ \ \ \ \ \ \ \ {\tt fun f x1 x2 $\cdots$ xn = t} \\
is interpreted as the following term: \\ 
\ \ \ \ \ \ \ \ {\tt f = fn x1 => (fn x2 => ( $\cdots$ (fn xn => t) $\cdots$ ))} \\
We only consider closed term (or combinator) \\
\ \ \ \ \ \ \ \ {\tt $\vdash$ t:A}. \\
The following proposition is proved easily by structural induction: 
\begin{proposition}
\label{propSecTypingRules-1}
If \mbox{{\tt x1:A1,...,xn:An|-t:B} then} \\
\ \ \ \ \ \ \ \ {\tt fun f x1 x2 $\cdots$ xn = t} \\
is a well formed function declaration of Standard ML. 
\end{proposition}

\paragraph{Term Reduction Rules}
Two of our reduction rules are\\
\ \ \ \ \ \ \ \ ($\beta$): {\tt (fn x=>t)s} \ \ $\Rightarrow$ \ \  {\tt t[s/x]} \\
\ \ \ \ \ \ \ \ ($\TENS$-red): {\tt let val (x,y)=(u,v) in w end} \ \ $\Rightarrow$ \ \  {\tt w[u/x,v/y]} \\
In fact, in Standard ML, {\tt s,u,v} must be {\it values} in order for these rules to be applied. 
But Linear Lambda Calculus satisfies SN and CR properties. So we don't need to care the evaluation order. 
Then note that if a function {\tt f} is defined by \\ 
\ \ \ \ \ \ \ \ {\tt fun f x1 x2 $\cdots$ xn = t}\\ 
and \\
\ \ \ \ \ \ \ \ {\tt x1:A1,...,xn:An|-s:B}, $\, \, \, \,$ {\tt |-t1:A1}, $\, \,$ $\ldots$, $\, \, \, \, $ {\tt |-tn:An} \\
then, we have \\
\ \ \ \ \ \ \ \ {\tt f t1 $\cdots$ tn} \ \ $\Rightarrow$ \ \ {\tt t[t1/x1,$\ldots$,tn/xn]} .\\
Moreover we need the following reduction for a theoretical reason, which is absent from Standard ML:\\
\ \ \ \ \ \ \ \ ($\eta$): {\tt t} \ \ $\Rightarrow$ \ \  {\tt (fn x => t x)} \\
In the following $=_{\beta \eta}$ denotes the congruence relation generated by the three reduction rules.

\section{Review of Mairson's Proof}
In this section we review the proof in \cite{Mai04} briefly.
Below by normal forms we mean $\beta \eta$-long normal forms. 
The basic construct is the following term:\\
{\tt - fun Pair x y z = z x y;} \\
{\scriptsize {\tt val Pair = fn : 'a -> 'b -> ('a -> 'b -> 'c) -> 'c}} \\
Using this, we define {\tt True} and {\tt False}: \\
{\tt - fun True x = Pair x y;} \\
{\scriptsize {\tt val True = fn : 'a -> 'b -> ('a -> 'b -> 'c) -> 'c}} \\
{\tt - fun False x = Pair y x;} \\
{\scriptsize {\tt val True = fn : 'a -> 'b -> ('b -> 'a -> 'c) -> 'c}} \\
Note that these are the normal forms of $\MBB{B}_{\rm MH}$. 
In order to define the term {\tt Copy} two auxiliary terms are needed: \\
{\tt - fun I x = x;} \\
{\scriptsize {\tt val I = fn : 'a -> 'a}} \\
{\tt - fun id B = B I I I ;} \\
{\scriptsize {\tt val id = fn : (('a -> 'a) -> ('b -> 'b) -> ('c -> 'c) -> 'd) -> 'd}} \\
The formal argument {\tt B} is supposed to receive {\tt True} or {\tt False}. 
It is easy to see that
\[
{\tt id} \, \, \, \, {\tt True} \, \, \Rightarrow^{\ast} \, \, {\tt I}, 
\quad \quad
{\tt id} \, \, \, \, {\tt False} \, \, \Rightarrow^{\ast} \, \, {\tt I}, 
\]
Then the term {\tt Copy} is defined as follows:\\
{\tt - fun Copy P = P (Pair True True) (Pair False False)} \\
\ \ \ \ \ \ {\tt (fn U => fn V => } \\
\ \ \ \ \ \ \ \ \ \ \ {\tt U (fn u1 => fn u2 => } \\
\ \ \ \ \ \ \ \ \ \ \ \ \ \ \ \ {\tt V (fn v1 => fn v2 => } \\
\ \ \ \ \ \ \ \ \ \ \ \ \ \ \ \ \ \ \ \ \
{\tt ((id v1) u1, (id v2) u2)))))} \\
We omit its type since it is too long. 
The formal argument {\tt P} is supposed to receive {\tt True} or {\tt False}. 
While \cite{Mai04} uses continuation passing style, 
the above term not since we have the $\TENS (= \ast)$-connective and can do a direct encoding using this connective.  
Then \\
{\tt Copy True;} \\
{\scriptsize {\tt val it=(fn,fn):('a -> 'b -> ('a -> 'b -> 'c) -> 'c)*('d -> 'e -> ('d -> 'e -> 'f) -> 'f)}} \\
{\tt Copy False;} \\
{\scriptsize {\tt val it=(fn,fn):('a -> 'b -> ('b -> 'a -> 'c) -> 'c)*('d -> 'e -> ('e -> 'd -> 'f) -> 'f)}} \\
(in fact, since SML/NJ does not allow any function values, it gives warnings, but the results are basically the same).
These are {\tt (True, True)} and {\tt (False, False)} respectively since
\begin{eqnarray*}
{\tt Copy} \, \,  \, \, {\tt True} \, \, \Rightarrow^{\ast} \, \, (({\tt id} \, \,  \, \, {\tt False}) \, \, \, {\tt True}, \, \, \, \, ({\tt id} \, \,  \, \, {\tt False}) \, \, \, \, {\tt True}) \\
{\tt Copy} \, \,  \, \, {\tt False} \, \, \Rightarrow^{\ast} \, \, (({\tt id} \, \,  \, \, {\tt True}) \, \, \, \, {\tt False}, \, \, \, \, ({\tt id} \, \,  \, \, {\tt True}) \, \, \, \, {\tt False}) 
\end{eqnarray*}
The basic observation here is that 
\begin{itemize}
\item the type of {\tt P} is unifiable with that of both {\tt True} and {\tt False};
\item the types of {\tt Copy True} and {\tt Copy False} are desirable ones, i.e.,
both have $\MBB{B}_{\rm MH} \TENS \MBB{B}_{\rm MH}$ as a instance. 
\end{itemize}
Since the {\it and} gate can be defined similarly and more easily and 
the {\it not} gate without any ML-polymorphism, it is concluded that all the boolean gates can be defined over $\MBB{B}_{\rm MH}$.

\section{A Partial Solution}
In this section we present our failed attempt.\\
Let the following two terms be {\tt True'} and {\tt False'}: \\
{\tt fun True' x y f = Pair x (f y);} \\
{\scriptsize {\tt val True' = fn : 'a -> 'b -> ('b -> 'c) -> ('a -> 'c -> 'd) -> 'd}} \\
{\tt fun False' x y f = Pair (f x) y;} \\
{\scriptsize {\tt val False' = fn : 'a -> 'b -> ('a -> 'c) -> ('c -> 'b -> 'd) -> 'd}} \\
Both are normal forms of $\MBB{B}_{\rm red} = p \LIMP p \LIMP (p \LIMP p) \LIMP (p \LIMP p \LIMP p) \LIMP p$, which have the six normal forms. 
Below we define a copy function for them as in the previous section. 
In order to do that, we need several auxiliary terms: \\
{\tt fun not' f x y g h = f y x g (fn u => fn v => (h v u));} \\
{\scriptsize {\tt val not' = fn}} \\
\ \ \ \ \ {\scriptsize {\tt  : ('a -> 'b -> 'c -> ('d -> 'e -> 'f) -> 'g) }} \\
\ \ \ \ \ \ \ \ \ \ \ \ \ \ \ {\scriptsize {\tt    -> 'b -> 'a -> 'c -> ('e -> 'd -> 'f) -> 'g}} \\
The term {\tt not'} is the {\it not} gate for the new boolean values. \\
{\tt fun swap f g = f (fn u => fn v => g v u);}\\
{\scriptsize {\tt val swap = fn : (('a -> 'b -> 'c) -> 'd) -> ('b -> 'a -> 'c) -> 'd}}\\
We note that
\begin{eqnarray*}
{\tt swap} \, \, ({\tt Pair} \, \, \, \, {\tt False'} \, \, \, \, {\tt True'}) \, \, {\tt g} 
& =_{\beta \eta} & 
({\tt Pair} \, \, \, \, {\tt False'} \, \, \, \, {\tt True'}) ({\tt fn} \, \, {\tt u} \, \, {\tt => } \, \, {\tt fn} \, \, {\tt v} \, \, {\tt =>} \, \, {\tt g} \, \, \, \, {\tt u} \, \, \, \, {\tt v}) \\
& =_{\beta \eta} & {\tt g} \, \, \, \, {\tt True'} \, \, \, \, {\tt False'} 
\, \,  =_{\beta \eta} \, \, {\tt Pair} \, \, \, \, {\tt True'} \, \, \, \, {\tt False'} \, \, \, \, {\tt g}
\end{eqnarray*}
The term {\tt newid} is similar to {\tt id}, but receives four arguments: \\
{\tt fun newid B' = B' I I I I;} \\
{\scriptsize {\tt val newid = fn}} \\
\ \ \ \ \ {\scriptsize {\tt  : (('a -> 'a) -> ('b -> 'b) -> ('c -> 'c) -> ('d -> 'd) -> 'e) -> 'e}} \\
The term {\tt constNot} is also similar to {\tt id}, but always returns {\tt not'}: \\
{\tt fun constNot B' = B' I not' I I;}\\
{\scriptsize {\tt val constNot = fn}} \\
\ \ \ \ \ {\scriptsize {\tt  : (('a -> 'a) }} \\
\ \ \ \ \ \ \ \ \ \ {\scriptsize {\tt     -> (('b -> 'c -> 'd -> ('e -> 'f -> 'g) -> 'h)}} \\
\ \ \ \ \ \ \ \ \ \ {\scriptsize {\tt         -> 'c -> 'b -> 'd -> ('f -> 'e -> 'g) -> 'h)}} \\
\ \ \ \ \ \ \ \ \ \ {\scriptsize {\tt        -> ('i -> 'i) -> ('j -> 'j) -> 'k)}} \\
\ \ \ \ \ \ \ \ \ \ {\scriptsize {\tt    -> 'k}} \\
The formal argument {\tt B'} in {\tt newid} and {\tt constNot} is supposed to receive {\tt True'} and {\tt False'}. 
We can easily see
\begin{eqnarray*}
{\tt newid} \, \, \, \, {\tt True'} \, \, \Rightarrow^{\ast} \, \, {\tt I}, &
\quad \quad 
{\tt newid} \, \, \, \, {\tt False'} \, \, \Rightarrow^{\ast} \, \, {\tt I}, \\
{\tt constNot} \, \, \, \, {\tt True'} \, \, \Rightarrow^{\ast} \, \, {\tt not}, &
\quad \quad 
{\tt constNot} \, \, \, \, {\tt False'} \, \, \Rightarrow^{\ast} \, \, {\tt not}
\end{eqnarray*}
Under the preparation above, we can define {\tt Copy'} as follows: \\
{\tt fun Copy' P' = P' (Pair False' True') (Pair False' True') swap} \\
\ \ \ \ \ \ \ \ \ \ {\tt (fn U => fn V => } \\
\ \ \ \ \ \ \ \ \ \ \ \ \ \ \ \ \ \ \ \ {\tt U (fn u1 => fn u2 => } \\
\ \ \ \ \ \ \ \ \ \ \ \ \ \ \ \ \ \ \ \ \ \ \ \ \ \ \ \  {\tt V (fn v1 => fn v2 => } \\
\ \ \ \ \ \ \ \ \ \ \ \ \ \ \ \ \ \ \ \ \ \ \ \ \ \ \ \ \ \ \ \ \ \ \ \ \ {\tt ((constNot v1) u1,  (newid v2) u2))));}\\
Again we omit the type. 
The formal parameter {\tt P} is supposed to receive {\tt True'} or {\tt False'}. 
Then
\begin{eqnarray*}
{\tt Copy'} \, \, \, \, {\tt True'} & \Rightarrow^{\ast} &
(({\tt constNot} \, \, \,  {\tt True}) \, \, \, \, {\tt False},  \, \, ({\tt newid} \, \, \, \,  {\tt False}) \, \, \, \, {\tt True}) \\
& \Rightarrow^{\ast} & ({\tt True}, \, \, {\tt True}) \\
{\tt Copy'} \, \, \, \, {\tt False'} & \Rightarrow^{\ast} &
(({\tt constNot} \, \, \,  {\tt False}) \, \, \, \, {\tt True},  \, \, ({\tt newid} \, \, \, \,  {\tt True}) \, \, \, \, {\tt False}) \\
& \Rightarrow^{\ast} & ({\tt False}, \, \, {\tt False}) 
\end{eqnarray*}
Unfortunately we could not find a term that represents the {\it and} gate over {\tt True'} and {\tt False'}.
So, we must find a similar, but different substitute. 
Fortunately we have found a solution described in the next section.

\section{Our Solution}
Our solution uses the type $\MBB{B} = (p \LIMP p) \LIMP (p \LIMP p) \TENS (p \LIMP p)$, 
which has the two normal forms:\\
{\tt fun True'' x = (fn z => z, fn y => x y);}\\
{\scriptsize {\tt val True'' = fn : ('a -> 'b) -> ('c -> 'c) * ('a -> 'b)}}\\
{\tt fun False'' x  = (fn y => x y,  fn z => z);}\\
{\scriptsize {\tt val False'' = fn : ('a -> 'b) -> ('a -> 'b) * ('c -> 'c)}}\\
The linear distributive transformation (which was called third-order reduction in \cite{Mat07})
turn $\MBB{B}$ into 
$\MBB{B}_{\rm red}$.
The next term is its internalized version: \\
{\tt fun LDTr h x y f z} \\
\ \ \ \ \ \ \ \ \ \ {\tt = let val (k, l) = h f in z (k x) (l y) end;}\\
{\scriptsize {\tt val LDTr = fn}} \\
\ \ \ \ \ \  {\scriptsize {\tt : ('a -> ('b -> 'c) * ('d -> 'e))}}\\
\ \ \ \ \ \ \ \ \ {\scriptsize {\tt -> 'b -> 'd -> 'a -> ('c -> 'e -> 'f) -> 'f}}\\
Then 
\[
{\tt LDTr} \, \, \, \, {\tt True''} \, \, =_{\beta \eta} \, \, {\tt True'}, 
\quad \quad 
{\tt LDTr} \, \, \, \, {\tt False''} \, \, =_{\beta \eta} \, \, {\tt False'} 
\]
Our {\it not} gate for {\tt True''} and {\tt False''} is \\
{\tt fun not'' h f = let val (k, l) = h f in (l, k) end;} \\
{\scriptsize {\tt val not'' = fn : ('a -> 'b * 'c) -> 'a -> 'c * 'b}} \\
Then
\[
{\tt not''} \, \, \, {\tt True''} \, \, =_{\beta \eta} \, \, {\tt False''},
\quad \quad 
{\tt not''} \, \, \, {\tt False''} \, \, =_{\beta \eta} \, \, {\tt True''}
\]
Moreover we can write down a {\it and} gate for them as follows:\\
{\tt fun and'' f g h = let val (u, v) = g (fn k => h k) in } \\
\ \ \ \ \ \ \ \ \ \ {\tt (let val (x, y) = f (fn w => v w) in}\\
\ \ \ \ \ \ \ \ \ \ \ \ \ \ \ \ {\tt (fn s => x (u s), fn t => y t) end) end;}\\
{\scriptsize {\tt val and'' = fn}} \\
\ \ \ \ \ \ {\scriptsize {\tt  : (('a -> 'b) -> ('c -> 'd) * ('e -> 'f))}} \\
\ \ \ \ \ \ \ \ \ \ {\scriptsize {\tt -> (('g -> 'h) -> ('i -> 'c) * ('a -> 'b))}} \\
\ \ \ \ \ \ \ \ \ \ \ \ \ \ {\scriptsize {\tt -> ('g -> 'h) -> ('i -> 'd) * ('e -> 'f)}}\\
Note that the definition of {\tt and''} does not use any ML-polymorphism. Then 
\begin{eqnarray*}
{\tt and''} \, \, \, {\tt True''} \, \, \, {\tt True''} \, \, =_{\beta \eta} \, \, {\tt True''},
& 
{\tt and''} \, \, \, {\tt False''} \, \, \, {\tt False''} \, \, =_{\beta \eta} \, \, {\tt False''}, \\
{\tt and''} \, \, \, {\tt True''} \, \, \, {\tt False''} \, \, =_{\beta \eta} \, \, {\tt False''},
& 
{\tt and''} \, \, \, {\tt False''} \, \, \, {\tt True''} \, \, =_{\beta \eta} \, \, {\tt False''}
\end{eqnarray*}
Next, we define a copy function for {\tt True''} and {\tt False''}. 
In order to do that, we need a modified version of {\tt constNot}:\\
{\tt fun constNot'' B'' = B'' I not'' I I;}\\
Then we can easily see
\begin{eqnarray*}
{\tt newid} \, \, \, \, {\tt True''} \, \, \Rightarrow^{\ast} \, \, {\tt I}, &
\quad \quad 
{\tt newid} \, \, \, \, {\tt False''} \, \, \Rightarrow^{\ast} \, \, {\tt I}, \\
{\tt constNot''} \, \, \, \, {\tt True''} \, \, \Rightarrow^{\ast} \, \, {\tt not}, &
\quad \quad 
{\tt constNot''} \, \, \, \, {\tt False''} \, \, \Rightarrow^{\ast} \, \, {\tt not}
\end{eqnarray*}
Under the preparation above, we can define {\tt Copy''}, which is a modified version
of {\tt Copy'} as follows: \\
{\tt fun Copy'' P} \\
\ \ \ \ \ \ \ {\tt = LDTr P (Pair False'' True'') (Pair False'' True'') swap}\\
\ \ \ \ \ \ \ \ \ \ {\tt (fn U => fn V => }\\
\ \ \ \ \ \ \ \ \ \ \ \ \ \ {\tt U (fn u1 => fn u2 => } \\
\ \ \ \ \ \ \ \ \ \ \ \ \ \ \ \ \ \  {\tt V (fn v1 => fn v2 => }\\
\ \ \ \ \ \ \ \ \ \ \ \ \ \ \ \ \ \ \ \ \ \ {\tt ((constNot'' (LDTr v1)) u1,  (newid (LDTr v2)) u2))));}\\
Then 
\[
{\tt Copy''} \, \, \, \, {\tt True''} \, \, \Rightarrow^{\ast} \, \, ({\tt True''}, \, \, {\tt True''}), \quad \quad 
{\tt Copy''} \, \, \, \, {\tt False''} \, \, \Rightarrow^{\ast} \, \, ({\tt False''}, \, \, {\tt False''}) 
\]
From what precedes we can conclude that 
we can represent all the boolean gates over $\MBB{B}$.

\section{Concluding Remarks}
In this paper we showed that $\MBB{B}_{\rm MH}$ is not the only choice
in order to establish P-time completeness of Linear Lambda Calculus. 
We note that we found the term {\tt and''} manually using proof nets syntax (and then translating 
the proof net into {\tt and''}), 
but {\tt Copy'} and {\tt Copy''} interactively with Standard ML of New Jersey.
\\
From our result a natural question comes up: 
which linear type other than $\MBB{B}_{\rm MH}$ and $\MBB{B}$ and its two normal forms establishes P-time completeness of Linear Lambda Calculus?
For example it is unlikely that $\MBB{B}' = p \LIMP (p \LIMP p) \LIMP (p \LIMP p) \LIMP p$
and its two normal forms establish that. 
But it is an easy exercise to show that $(p \TENS p) \LIMP (p \TENS p)$ and its two normal forms can do that. \\
As wrote before, we could not prove that $\MBB{B}_{\rm red}$ and its normal forms
{\tt True'} and {\tt False'} establish P-time completeness of Linear Lambda Calculus.
But we also could not prove that they cannot establish that. 
At this moment we do not have any idea to do that. 
Our type $\MBB{B}$ and its generalization
$(p \LIMP p) \LIMP \overbrace{(p \LIMP p) \TENS \cdots \TENS (p \LIMP p)}^{n}$ have
further interesting properties. 
For example we can establish weak typed B\"{o}hm theorem over $\MBB{B}$. 
But the subject is beyond the scope of this paper, and will be discussed elsewhere.

\end{document}